\documentclass{article}

\usepackage{arxiv}

\usepackage[utf8]{inputenc} %
\usepackage[T1]{fontenc}    %
\usepackage{hyperref}       %
\usepackage{url}            %
\usepackage{booktabs}       %
\usepackage{amsfonts}       %
\usepackage{nicefrac}       %
\usepackage{microtype}      %
\usepackage{color}
\usepackage[english]{babel}
\usepackage{csquotes}
\usepackage{amsmath}
\usepackage{amsthm}
\usepackage{amssymb}
\usepackage{stmaryrd}
\usepackage{graphicx}
\usepackage[style=authoryear,natbib=true,maxcitenames=2,maxbibnames=10,
uniquelist=false,backend=biber]{biblatex}
\bibliography{mrrce}

\newtheorem{prop}{\textbf{Proposition}}

\providecommand{\tabularnewline}{\\}
\usepackage{algorithm,algpseudocode}
\usepackage{bbm}

\title{Capturing Between-Tasks Covariance and Similarities\\
Using Multivariate Linear Mixed Models}

\author{
  Aviv Navon \\
  Department of Statistics and Operations Research\\
  Tel-Aviv university\\
  Tel-Aviv, Israel \\
  \texttt{avivnavon@mail.tau.ac.il} \\
   \And
 Saharon Rosset\\
  Department of Statistics and Operations Research\\
  Tel-Aviv university\\
  Tel-Aviv, Israel \\
  \texttt{saharon@tauex.tau.ac.il} \\
}

\begin{document}
\maketitle

\begin{abstract}
We consider the problem of predicting several response variables using
the same set of explanatory variables. This setting naturally induces
a group structure over the coefficient matrix, in which every explanatory
variable corresponds to a set of related coefficients. Most of the
existing methods that utilize this group formation assume that the
similarities between related coefficients arise solely through a joint
sparsity structure. In this paper, we propose a procedure for constructing
an estimator of a multivariate regression coefficient matrix that
directly models and captures the within-group similarities, by employing
a multivariate linear mixed model formulation, with a joint estimation
of covariance matrices for coefficients and errors via penalized likelihood.
Our approach, which we term Multivariate random Regression with Covariance
Estimation (MrRCE) encourages structured similarity in parameters,
in which coefficients for the same variable in related tasks sharing
the same sign and similar magnitude. We illustrate the benefits of
our approach in synthetic and real examples, and show that the proposed
method outperforms natural competitors and alternative estimators
under several model settings.
\end{abstract}

\keywords{Covariance selection \and EM algorithm \and Multivariate
regression \and Penalized likelihood \and Regularization methods \and Sparse precision
matrix}

\section{Introduction}

In many cases, a common set of predictor variables is used for predicting
different but related target variables. For example, an on-demand
transportation company may attempt forecasting demand and supply in
different time frames and geographic locations; a real-estate firm
may be interested in predicting both the construction costs and the
sale prices of residential apartments, given a set of project's physical
and financial covariates, and external economic variables.

The general task of modeling multiple responses using a joint set
of covariates can be expressed using multivariate regression (MR),
or multiple response regression \textemdash{} a generalization of
the classical regression model to regressing $q>1$ responses on $p$
predictors. In the MR settings, one is presented with $n$ independent
observations, $\left\{ \left(X_{i},Y_{i}\right)\right\} _{i=1}^{n}$,
where $X_{i}\in\mathbb{R}^{p}$ and $Y_{i}\in\mathbb{R}^{q}$ contain
the predictors and responses for the $i$th sample, respectively.
Let $X=\left(X_{1},...,X_{n}\right)^{T}=\left(\mathbf{x}_{1},...,\mathbf{x}_{p}\right)\in\mathbb{R}^{n\times p}$
denote the predictor matrix and $Y=\left(Y_{1},...,Y_{n}\right)^{T}=\left(\mathbf{y}_{1},...,\mathbf{y}_{q}\right)\in\mathbb{R}^{n\times q}$
denote the response matrix. For simplicity of notation, assume that
the columns of $X$ and $Y$ have been centered so that we need not
consider an intercept term. We further assume that the i.i.d $N_{q}\left(0,\Sigma\right)$
error terms are collected into an $n\times q$ error matrix $E$,
where $\Sigma$ is the among-tasks covariance matrix. The multivariate
regression model is given by, 
\begin{equation}
Y=XB+E\label{eq:1.1}
\end{equation}
where $B$ is a $p\times q$ regression coefficient matrix. The random
matrices in (\ref{eq:1.1}) are assumed to follow a matrix-variate
normal distribution \citep{dawid1981some,gupta2018matrix}, $E\sim MVN_{n\times q}\left(0,I_{n},\Sigma\right)$
and $Y\sim MVN_{n\times q}\left(XB,I_{n},\Sigma\right)$. For reasons
that will later become clear, when considering the noise structure
of the MR model, the precision matrix, $\Omega=\Sigma^{-1}$, is commonly
the preferred object.

Straightforward prediction and estimation with the MR model can become
quite challenging when the number of predictors and responses is large
relative to $n$, as it requires one to estimate $pq$ parameters.
The univariate regression model ($q=1$) has been widely studied,
and numerous methods have been developed for variable selection (support
recovery) and coefficients estimation. A naive approach to the MR
problem is to apply one of these methods to each of the $q$ tasks
independently. However, in many cases, the different problems are
related, and this oversimplified approach fails to utilize all the
information contained in the data (see, e.g., \citet{breiman1997predicting, rothman2010sparse}).
For a review of Bayesian approaches for estimation and prediction
with the MR model see \citet{deshpande2017simultaneous} and references
therein.

In the MR literature, many approaches seek to reduce the number of
parameters to be estimated through a penalized (or constrained) least
squares framework. \citet{bunea2011optimal} generalized the classical
Reduced-Rank Regression (RRR) \citep{anderson1951estimating,izenman1975reduced,velu2013multivariate}
to high dimensional settings, estimating a low-rank coefficient matrix
by penalizing the rank of $B$. \citet{yuan2007dimension} proposed
a method called Factor Estimation and Selection (FES), in which an
$L_{1}$-penalty is applied to the singular values of $B$. FES induces
sparsity in the singular values of $B$, conducting dimension reduction
and coefficients estimation simultaneously. One major drawback of
dimension reduction techniques, is that the interpretation of the
model is often limited, in terms of the original data, since the set
of predictors is reduced to a few important principal factors.

\begin{figure}
	\centering
	\makebox{\includegraphics[scale=0.5]{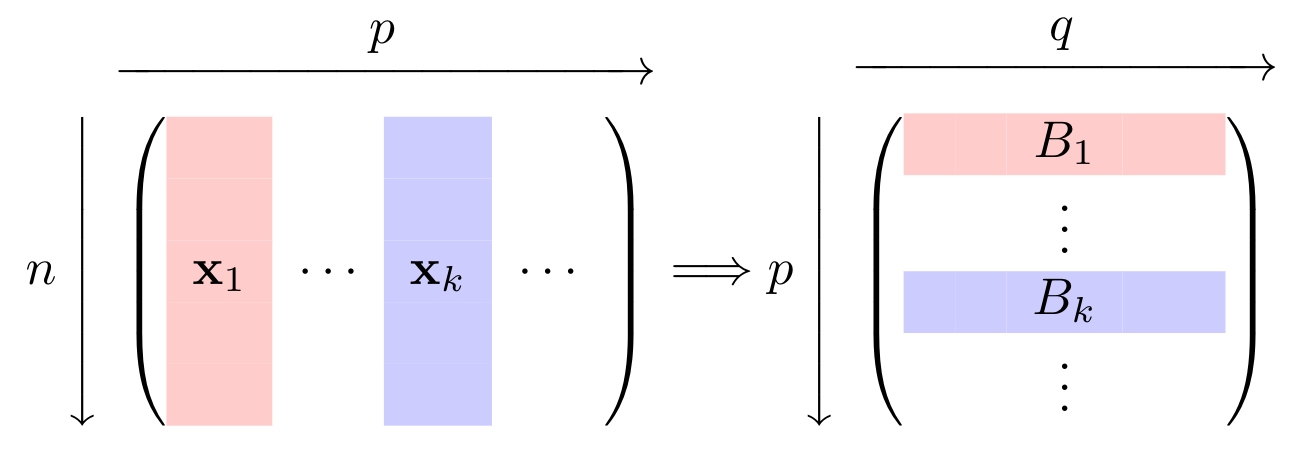}}
	\caption{\label{fig:mr-groups}The multivariate regression framework naturally
		induces a group structure over the coefficient matrix $B$, in which
		every explanatory variable, $\mathbf{x}_{i}$, corresponds to a group
		of $q$ coefficients $B_{i}=\left(\beta_{i1},...,\beta_{iq}\right)^{T}$.}
\end{figure}

The multivariate regression framework naturally induces a group structure
over the coefficient matrix, $B$, in which every explanatory variable,
$\mathbf{x}_{i}\text{ for }i=1,...,p$, corresponds to a group of
$q$ coefficients, $B_{i}=\left(\beta_{i1},...,\beta_{iq}\right)$
(see Figure \ref{fig:mr-groups}). While many approaches make no assumption
over the group structure, others utilize it for learning structured
sparsity. In the multi-task learning literature, the $L_{1}/L_{2}$-penalty,
also known as the group lasso penalty \citep{yuan2006model}, has
been applied with the rows of $B$ as groups. The $L_{1}/L_{2}$-penalty
can be viewed as an intermediate between the $L_{1}$-penalty used
in lasso regression \citep{tibshirani1996regression} and the $L_{2}$-penalty
used in ridge regression \citep{hoerl1970ridge}, aimed at utilizing
the relatedness among tasks for identifying the joint support, i.e.,
the set of predictors with non-zero coefficients across all $q$ responses
\citep{obozinski2009high}. \citet{peng2010regularized} proposed
a mixed constraint function, by applying both the lasso and the group
lasso penalties to the elements and rows of $B$, respectively. This
approach produces element-wise as well as row-wise sparsity in the
coefficient matrix. \citet{turlach2005simultaneous} studied a different
constraint function, placing an $L_{\infty}$-penalty over the rows
of $B$. As noted by the authors, this method is only suitable for
variable selection and not for estimation. Extensions of mixed norm
penalties to overlapping groups have been proposed in order to handle
more general and complex group structures (see, e.g., \citet{kim2012tree,li2015multivariate}).
These methods produce highly interpretable models, however, they are
limited to the case $\Omega\propto I_{n}$, and do not account for
correlated errors. \citet{rothman2010sparse,chen2016sparse,wilms2018algorithm}
have recently shown that accounting for this additional information
in MR problems can be beneficial for both coefficients estimation
and prediction.

In multivariate normal theory, the entries of $\Omega$ that equal
zero correspond to pairs of variables that are conditionally independent,
given all of the other variables in the data. The problem of sparse
precision matrix estimation has drawn considerable recent attention,
and several methods have been proposed for both support recovery and
parameter estimation. Perhaps the most widely used approach is the
graphical lasso \citep{friedman2008sparse}, in which simultaneous
sparsity structure identification and coefficients estimation are
achieved by minimizing the $L_{1}$-regularized negative log-likelihood
function of $\Omega$ \citep{yuan2007model,banerjee2008model,rothman2008sparse}.
Recently, sparse precision matrix estimation has also been considered
in regression frameworks, in which the main goal for this explicit
estimation is to improve prediction \citep{witten2009covariance,rothman2010sparse}. 

\citet{rothman2010sparse} proposed Multivariate Regression with Covariance
Estimation (MRCE), a method for sparse multivariate regression that
directly accounts for correlated errors. MRCE minimizes the negative
log-likelihood function with an $L_{1}$-penalty for both $B$ and
$\Omega$,
\begin{equation}
\arg\min_{B,\Omega}-n\log\left|\Omega\right|+\text{tr}\left[\frac{1}{n}\left(Y-XB\right)^{T}\Omega\left(Y-XB\right)\right]+\lambda_{1}\left\Vert B\right\Vert _{1}+\lambda_{2}\sum_{j\neq j'}\left|\omega_{jj'}\right|\label{eq:1.3}
\end{equation}
where $\text{tr}\left(\cdot\right)$ denotes the trace, $\lambda_{1}$
and $\lambda_{2}$ are the regularization parameters and $\omega_{jj'}$
is the $\left(j,j'\right)$ element of $\Omega$. \citet{lee2012simultaneous}
extended the approach of \citet{rothman2010sparse} to allow for weighted
$L_{1}$-penalties over the elements of $B$ and $\Omega$. \citet{yin2011sparse}
considered a similar objective to the one in (\ref{eq:1.3}), and
proposed an algorithm for the sparse estimation of the coefficient
and inverse covariance matrices. However, unlike \citet{rothman2010sparse},
their method aimed at improving the estimation of $\Omega$, rather
than $B$. Our work further leverages correlations between the different
problems to improve the accuracy of the estimators and predictions,
by not only accounting for the correlation between the error terms
but the similarities between the coefficients as well.

While MRCE accounts for correlated responses through the precision
matrix $\Omega$, it does not learn structured sparsity in $B$, essentially
selecting relevant covariates for each response separately. In a recent
work, \citet{wilms2018algorithm} proposed an algorithm for the multivariate
group lasso with covariance estimation, replacing the lasso penalty
in (\ref{eq:1.3}) with an $L_{1}/L_{2}$-penalty over a pre-specified
group structure. \citet{chen2016sparse} developed a method within
the reduced-rank regression framework that simultaneously performs
variable selection and sparse precision matrix estimation. These methods
for learning group sparsity assume that the sparsity structure is
known a-priori. Instead, \citet{sohn2012joint} proposed an approach
for group sparse multivariate regression that can jointly learn both
the response structure and regression coefficients with structured
sparsity.

All the above methods which considered a group structure over the
coefficient matrix, essentially assume that the within-group similarities
arise solely through a joint sparsity structure. In many applications,
these structured (and unstructured) sparsity assumptions are not suitable,
for instance, if one expects many covariates of small or medium effect.
Furthermore, these sparse estimators encourage within-group coefficients
to be of similar absolute magnitude, and do not favor same sign coefficients.
However, in various real-life examples it is more natural to encourage
coefficients within the same group to also share a sign. To address
these issues, we construct an estimator for the multivariate regression
by directly modeling and capturing the within-group similarities,
while also accounting for the error covariance structures. Our method,
titled Multivariate random Regression with Covariance Estimation (MrRCE),
involves a multivariate linear mixed model with an underlying group
structure over the coefficient matrix, designed to encourage related
coefficients to share a common sign and similar magnitude.

Multivariate Linear Mixed Models (mvLMMs) \citep{henderson1984applications}
are MR models that relate a joint set of covariates to multiple correlated
responses. mvLMMs are applied in many real-life problems and frequently
used in genetics due to their ability to account for relatedness among
observations (see, e.g., \citet{kruuk2004estimating,kang2010variance,korte2012mixed,vattikuti2012heritability}).
The mvLMMs model can be viewed as a generalization of MR (similar
to the way Linear Mixed Models (LMMs) are a generalization of linear
regression models), allowing both fixed and random effects. Consider
the MR problem (\ref{eq:1.1}), but with an additional term for the
set of random predictors, collected into the matrix $Z=\left(Z_{1},...,Z_{n}\right)^{T}=\left(\mathbf{z}_{1},...,\mathbf{z}_{r}\right)\in\mathbb{R}^{n\times r}$.
The mvLMM model is given by,
\begin{align}
Y & =XB+Z\Gamma+E\label{eq:1.4}\\
E & \sim MVN_{n\times q}\left(0,I_{n},\Sigma\right)\text{, }\Gamma\sim MVN_{r\times q}\left(0,R,G\right)\nonumber 
\end{align}
where $B$ is a $p\times q$ fixed effect coefficient matrix and $\Gamma$
is an $r\times q$ random effect coefficient matrix. Here, $R$ and
$G$ are the common covariance matrices of columns and rows of $\Gamma$,
respectively.

In this paper we consider the problem of estimation and prediction
under the multivariate random effect regression \textemdash{} an mvLMMs
model strictly involving random effects, 
\begin{equation}
Y=Z\Gamma+E\label{eq:1.5}
\end{equation}
Under the proposed formulation and unlike the standard mvLMM framework,
we are interested in estimating not only the covariance components
but also in predicting the random component $\Gamma$. Our method
accounts for correlations between responses and similarities among
coefficients, captured by estimating a joint equicorrelation covariance
matrix for the rows of $\Gamma$ (see Eq. \ref{eq:2.1} for details).
Hence, the MrRCE method is an example of what one could call \textit{structured
	similarity} learning, in which the different coefficient groups are
assumed to be independent, whereas a within-group similarity is encouraged.
This covariance structure for the random coefficient matrix reduces
the MR problem of estimating $pq$ parameters, into the problem of
estimating two covariance components \textemdash{} the coefficients'
common variance, and the \textit{intra-group correlation coefficient},
or \textit{similarity level}. The estimation of the covariance structure
is achieved through a penalized likelihood, adding an $L_{1}$-penalty
over the off-diagonal entries of $\Omega=\Sigma^{-1}$.

The remainder of the paper is structured as follows. Section \ref{sec:sec2}
describes the MrRCE method and corresponding Expectation-Maximization
(EM) based computational algorithm. Section \ref{sec:sec3} establishes
a connection between the proposed method and the multivariate Ridge
estimator. Simulation studies are performed in Section \ref{sec:sec4}
to compare our method with competing estimators, and Section \ref{sec:sec5}
contains two real data applications of MrRCE. Section \ref{sec:sec6}
concludes with a brief discussion.

\section{\label{sec:sec2}The MrRCE Method}

Consider the random effect regression model (\ref{eq:1.5}) with $r=p$.
Assume both the error matrix $E$ and the coefficient matrix $\Gamma$
follow a matrix variate normal distribution, 
\begin{align}
E & \sim MVN_{n\times q}\left(0,I_{n},\Sigma\right)\text{, }\Gamma\sim MVN_{p\times q}\left(0,I_{p},\sigma^{2}C\right)\label{eq:2.1}
\end{align}
Further assume an equicorrelation structure for the matrix $C$, controlled
by the unknown intra-group correlation coefficient $\rho\in[0,1)$,
\[
C=C_{\rho}=\begin{pmatrix}1 & \rho & \cdots & \rho\\
\rho & \ddots &  & \vdots\\
\vdots &  & \ddots & \rho\\
\rho & \cdots & \rho & 1
\end{pmatrix}
\]
The unknown parameter $\rho$ can be thought of as a relative measure
of the\textit{ within-group similarity} \citep{chatfield2010introduction}.
Large values for $\rho$ correspond to high similarity among members
of the same group, leading to a similar magnitude and same sign coefficients,
whereas $\rho=0$ corresponds to $i.i.d$ draws for the entries of
the coefficient matrix $\Gamma$.
We refer to the random variable $\Gamma$ as unobserved
data, and to $\left(Y,\Gamma\right)$ as the \textit{full
	data}. Denote the likelihood function of the full
data by $\mathcal{L}\left(\cdot\right)$, and the collection of parameters
by $\Theta=\left\{ \Omega,\sigma^{2},\rho\right\} $, we have,
\begin{align*}
\mathcal{L}\left(Y,\Gamma;\Theta\right) & =\mathcal{L}_{Y\mid\Gamma}\left(Y\mid\Gamma;\Theta\right)\mathcal{L}_{\Gamma}\left(\Gamma\mid\Theta\right)\\
& =\mathcal{L}_{Y\mid\Gamma}\left(Y\mid\Gamma;\Omega\right)\mathcal{L}_{\Gamma}\left(\Gamma\mid\sigma^{2},\rho\right)
\end{align*}
Thus, the negative log-likelihood function of the complete data is
given by (up to a constant),
\[
\ell\left(Y,\Gamma;\Theta\right)=\text{tr}\left[\frac{1}{n}\Omega\left(Y-Z\Gamma\right)^{T}\left(Y-Z\Gamma\right)\right]-\log\left|\Omega\right|+\text{tr}\left[\frac{1}{p}\Delta\Gamma^{T}\Gamma\right]-\log\left|\Delta\right|
\]
where $\Delta^{-1}=\sigma^{2}C$. We construct an estimator of $\Theta$
using a penalized normal log-likelihood, adding an $L_{1}$-penalty
over the off-diagonal entries of $\Omega$,

\begin{equation}
\hat{\Theta}=\arg\min_{\Theta}\ell\left(Y,\Gamma;\Theta\right)+\lambda_{\omega}\sum_{j\neq j'}\left|\omega_{jj'}\right|\label{eq:2.2}
\end{equation}
where $\lambda_{\omega}>0$ is a regularization parameter.

\subsection{The Algorithm}

We propose an iterative, EM-based \citep{dempster1977maximum}
algorithm for solving (\ref{eq:2.2}). Alg.~\ref{alg:the_alg} provides
a schematic overview of the MrRCE algorithm.

Using eigendecomposition (similar to \citet{zhou2014efficient,furlotte2015efficient}),
we write, 
\begin{equation}
C=UDU^{T}\text{ and }ZZ^{T}=LSL^{T}\label{eq:eigen}
\end{equation}
where $S$ and $D:=D_{\rho}=diag\left(d_{1}\left(\rho\right),...,d_{q}\left(\rho\right)\right)$
are diagonal matrices, and $U$ is independent of $\rho$. We then
multiply (\ref{eq:1.5}) by the orthogonal matrices $U\text{ and }L^{T}$
from the right and left correspondingly, to obtain, 
\[
\tilde{Y}=\tilde{Z}\tilde{\Gamma}+\tilde{E}
\]
where $\tilde{Y}=L^{T}YU$, $\tilde{Z}=L^{T}Z$, and,
\begin{align*}
\tilde{\Gamma} & =\Gamma U\sim MVN_{p\times q}\left(0,I_{p},\sigma^{2}U^{T}CU\right)=MVN_{p\times q}\left(0,I_{p},\sigma^{2}D_{\rho}\right)\\
\tilde{E} & =L^{T}EU\sim MVN_{n\times q}\left(0,L^{T}L=I_{n},\tilde{\Sigma}:=U^{T}\Sigma U\right)=MVN_{n\times q}\left(0,I_{n},\tilde{\Sigma}\right)
\end{align*}
We lose the $\tilde{\cdot}$ notation and assume (with a slight abuse
of notation) that the original data is of the form,
\begin{align}
& Y=Z\Gamma+E\label{eq:2.3}\\
& E\sim MVN_{n\times q}\left(0,I_{n},\Sigma:=\Omega^{-1}\right)\text{, }\Gamma\sim MVN_{p\times q}\left(0,I_{p},\sigma^{2}D_{\rho}\right)\nonumber 
\end{align}
namely,
\[
Y\sim MVN_{n\times q}\left(0,S,\sigma^{2}D_{\rho}\right)+MVN_{n\times q}\left(0,I_{n},\Sigma\right)
\]
Next, we describe an EM-based algorithm for solving (\ref{eq:2.2})
under the assumptions (\ref{eq:2.3}). \\
\\
\textbf{\textit{E-step}.} Denote
$\Theta_{t-1}$ the estimation for $\Theta$ at iteration $t-1$.
At step $t$, we wish to evaluate the following expressions,
\begin{align}
Q_{t}^{1}= & \mathbb{E}\left[\left(Y-Z\Gamma\right)^{T}\left(Y-Z\Gamma\right)\mid Y,\Theta_{t-1}\right]\label{eq:2.4}\\
Q_{t}^{2}= & \mathbb{E}\left[\Gamma^{T}\Gamma\mid Y,\Theta_{t-1}\right]\label{eq:2.5}
\end{align}
We let $\otimes$ denote the Kronecker product and $\text{vec}\left(\cdot\right)$
the vectorization operator\footnote{Let $\text{vec}\left(\cdot\right)$ denote the concatenation
	of a $k\times l$-dimensional matrix\textquoteright s columns into
	a $kl$-dimensional vector.}. For a matrix $A\in\mathbb{R}^{k\times p}$, we
let $A\Gamma:=G=\begin{pmatrix}\mathbf{g}_{1} & \cdots & \mathbf{g}_{q}\end{pmatrix}$,
with $\mathbf{g}_{j}$ the $j$th column of $G$. The joint distribution
of $\mathbf{g}=\text{vec}\left(G\right)$ and $\mathbf{y}=\text{vec}\left(Y\right)$
is given by,
\[
\begin{pmatrix}\mathbf{g}\\
\mathbf{y}
\end{pmatrix}\sim N\left(\mathbf{0},\begin{bmatrix}\Delta^{-1}\otimes AA^{T} & \Delta^{-1}\otimes AZ^{T}\\
\Delta^{-1}\otimes ZA^{T} & \Sigma\otimes I_{n}+\Delta^{-1}\otimes ZZ^{T}
\end{bmatrix}:=\begin{bmatrix}\Sigma_{11} & \Sigma_{12}\\
\Sigma_{21} & \Sigma_{22}
\end{bmatrix}\right)
\]
hence, the conditional distribution of $\mathbf{g}\mid\mathbf{y}$
is given by, 
\begin{equation}
\mathbf{g}\mid\mathbf{y}\sim N\left(\Sigma_{12}\Sigma_{22}^{-1}\mathbf{y},\Sigma_{11}-\Sigma_{12}\Sigma_{22}^{-1}\Sigma_{21}\right)\label{eq:2.6}
\end{equation}
In order to evaluate (\ref{eq:2.4}) and (\ref{eq:2.5}), we calculate
$\mathbb{E}\left[\Gamma\mid Y,\Theta_{t-1}\right]$ and $\mathbb{E}\left[\Gamma^{T}A^{T}A\Gamma\mid Y,\Theta_{t-1}\right]$
for $A=I_{p},Z$. The former is the Empirical-Best Linear Unbiased
Predictor (E-BLUP) \citep{henderson1975best,henderson1984applications}
(see \textbf{\textit{Predicting $\Gamma$}}
below), whereas the latter can be easily obtained from (\ref{eq:2.6})
since, 
\[
\mathbb{\mathbb{E}}\left[G^{T}G\mid Y,\Theta_{t-1}\right]_{i,j}=\mathbb{E}\left[\mathbf{g}_{i}^{T}\mathbf{g}_{j}\mid\mathbf{y},\Theta_{t-1}\right]
\]
\\
\textbf{\textit{M-step}.} The
minimization of the objective over $\Theta$ can be split into two
disjoint minimization problems:
\begin{align}
& \arg\min_{\Omega\succeq0}\text{tr}\left[\frac{1}{n}\Omega Q_{t}^{1}\right]-\log\left|\Omega\right|+\lambda_{\omega}\sum_{j\neq j'}\left|\omega_{jj'}\right|\label{eq:2.7}\\
& \arg\min_{\sigma>0,\rho\in[0,1)}\text{tr}\left[\frac{1}{p}\Delta Q_{t}^{2}\right]-\log\left|\Delta\right|\label{eq:2.8}
\end{align}
The first minimization problem is exactly the $L_{1}$-penalized precision
matrix estimation problem considered by \citet{yuan2007model,banerjee2008model,friedman2008sparse,rothman2010sparse,hsieh2011sparse},
among others. We solve (\ref{eq:2.7}) by applying the graphical lasso
algorithm of \citet{friedman2008sparse}. The second minimization
problem, (\ref{eq:2.8}), can be easily solved in closed-form by utilizing
the diagonal form of $\Delta$.\\
\\
\textbf{\textit{Predicting $\Gamma$}.} {Given
	$\hat{\Theta}$, our estimation for $\Theta$, we compute the E-BLUP
	\citep{henderson1975best,henderson1984applications} for $\boldsymbol{\gamma}=\text{vec}\left(\Gamma\right)$.
	Denote, $\tilde{Z}=I_{q}\otimes Z,\text{ }L=\hat{\sigma}^{2}\hat{D}_{\rho}\otimes I_{p}$
	and $R=\hat{\Omega}^{-1}\otimes I_{n}$, the E-BLUP $\mathbf{\boldsymbol{\gamma}}^{*}$
	for $\boldsymbol{\gamma}$, is given by, 
	\[
	\textbf{\ensuremath{\boldsymbol{\gamma}}}^{*}=\left(\tilde{Z}^{T}R^{-1}\tilde{Z}+L^{-1}\right){}^{-1}\tilde{Z}^{T}R^{-1}\mathbf{y}
	\]
	Alternatively, as proved by \citet{henderson1959estimation}, $\boldsymbol{\gamma}^{*}=L^{T}\tilde{Z}^{T}\Psi^{-1}\mathbf{y}$
	where, $\Psi=\tilde{Z}L\tilde{Z}^{T}+R$. In order to predict $\Gamma$,
	we simply compute $\Gamma^{*}=\text{unvec}\left(\boldsymbol{\gamma}^{*}\right)$,
	where $\text{unvec}\left(\cdot\right)$ represents the reversal of
	the $\text{vec}\left(\cdot\right)$ operation.}\\
\\
\textbf{\textit{Starting value and Stopping Criteria}.}
We initialize $\Omega_{0}=I_{q}$, $\Delta^{-1}=I_{q}$, and consider
two alternatives for the MrRCE algorithm's stopping criteria.
\begin{enumerate}
	\item Set a tolerance value, $\tau>0$. Iterate until
	the sum of absolute changes in the values of $\Theta$ in two successive
	iterations is smaller than the tolerance value.
	\item Set a tolerance value, $\tau>0$, and let $l_{t}$
	denote the log-likelihood at iteration $t$. Iterate until the relative
	change in the log-likelihood value, $\left|\frac{l_{t-1}-l_{t}}{l_{t-1}}\right|$,
	is smaller than $\tau$.
\end{enumerate}
\textbf{\textit{Convergence}.}
The MrRCE algorithm is a variant of the EM algorithm for penalized
likelihood, hence each step ensures a decrease in the objective, and
the algorithm's convergence is guaranteed (see e.g. \citealp{green1990use}).

\begin{algorithm}
	\begin{algorithmic}[1]
		\Require{Regularization parameter $\lambda_\omega > 0$.}
		\State \textbf{Initialize:} set $t=0$ and $\Omega_t=\Delta_t^-1=I_q$.
		\Repeat{}
		\vspace{.15cm} 
		\Statex{$t\leftarrow t+1$}
		\Statex{\textbf{\textit{E-step:}} calculate $Q_t^1=\mathbb{E}\left[\left(Y-Z\Gamma\right)^{T}\left(Y-Z\Gamma\right)\mid Y,\Theta_{t-1}\right]$}
		\Statex and $Q_t^2=\mathbb{E}\left[\Gamma^{T}\Gamma\mid Y,\Theta_{t-1}\right]$
		\Statex{\textbf{\textit{M-step:}} solve $\Omega_t=\arg\min_{\Omega\succeq0}\text{tr}\left[\frac{1}{n}\Omega Q_{t}^{1}\right]-\log\left|\Omega\right|+\lambda_{\omega}\sum_{j\neq j'}\left|\omega_{jj'}\right|$}
		
		\Statex{and $\left(\sigma_t, \rho_t\right)=\arg\min_{\sigma>0,\rho\in[0,1)}\text{tr}\left[\frac{1}{p}\Delta Q_{t}^{2}\right]-\log\left|\Delta\right|$}
		
		\vspace{.15cm} 
		\Until{stopping criterion is reached.}
		\State{\textbf{predict $\Gamma$:} compute the E-BLUP for $\Gamma$, $\Gamma^*=\text{unvec}\left(\mathbb{E}\left[\boldsymbol{\gamma} \mid \textbf{y},\Theta_t\right]\right)$.} 
		\State{\Return{$\left(\Gamma^*, \Theta_t \right)$}}
	\end{algorithmic}
	\caption{\textbf{(MrRCE):} EM-based optimization procedure (see text for details)}
	\label{alg:the_alg}
\end{algorithm}

\section{\label{sec:sec3}Connection to Ridge Regression}

We present a connection between the MrRCE method and the Ridge Regression
(RR) estimator \citep{hoerl1970ridge}. More specifically, we explore
a special case in which the BLUP for $\Gamma$ derived by the MrRCE
algorithm is equivalent to the multivariate RR estimator \citep{brown1980adaptive}.

Consider the model,
\begin{align*}
& \mathbf{y}=\tilde{Z}\boldsymbol{\gamma}+\boldsymbol{\epsilon}\\
& \boldsymbol{\epsilon}\sim N\left(\mathbf{0},\Sigma_{0}\otimes I_{n}:=\Sigma\right)\text{, }\boldsymbol{\gamma}\sim N\left(\mathbf{0},\Lambda_{0}\otimes I_{p}:=\Lambda\right)
\end{align*}
The joint distribution of $\left(\textbf{y},\boldsymbol{\gamma}\right)$
is given by,
\[
\left(\begin{array}{c}
\boldsymbol{\gamma}\\
\mathbf{y}
\end{array}\right)\sim N\left(\mathbf{0},\left[\begin{array}{cc}
\Lambda & \Lambda\tilde{Z}^{T}\\
\tilde{Z}\Lambda & \tilde{Z}\Lambda\tilde{Z}^{T}+\Sigma
\end{array}\right]\right)
\]
and the BLUP for the random coefficient vector is the expectation
of $\boldsymbol{\gamma}$ conditional on $\mathbf{y}$,
\begin{align*}
\hat{\boldsymbol{\gamma}}_{\text{BLUP}} & =\mathbb{E}\left[\boldsymbol{\gamma}\mid\mathbf{y}\right]\\
& =\Lambda\tilde{Z}^{T}\left(\tilde{Z}\Lambda\tilde{Z}^{T}+\Sigma\right)^{-1}\mathbf{y}
\end{align*}
The RR estimator can be extended to the multivariate case as in \citet{brown1980adaptive},
\[
\hat{\boldsymbol{\gamma}}_{RR}=\left(\tilde{Z}^{T}\tilde{Z}+K\right)^{-1}\tilde{Z}^{T}\mathbf{y}
\]
where $K\succ0$ is the $pq\times pq$ ridge matrix. We apply the
generalized Sherman-Morrison-Woodbury \citep{sherman1950adjustment,woodbury1950inverting}
formula to the inverse of $\tilde{Z}^{T}\tilde{Z}+K$, to obtain,
\begin{align}
\hat{\boldsymbol{\gamma}}_{\text{RR}} & =K^{-1}\tilde{Z}^{T}\left[I-\left(I+\tilde{Z}K^{-1}\tilde{Z}^{T}\right)^{-1}\tilde{Z}K^{-1}\tilde{Z}^{T}\right]\mathbf{y}\label{eq:3.1}
\end{align}
Eq. \ref{eq:3.1} can be simplified as follow, 
\begin{align*}
\hat{\boldsymbol{\gamma}}_{\text{RR}} & =K^{-1}\tilde{Z}^{T}\left[\tilde{Z}K^{-1}\tilde{Z}^{T}+I\right]^{-1}\mathbf{y}
\end{align*}
Thus, under the $i.i.d$ error model, i.e., $\Sigma_{0}=\sigma_{\epsilon}^{2}I_{q}$,
setting $K=\left(\Sigma_{0}\otimes I_{p}\right)\Lambda^{-1}$ yields,
\begin{align*}
\hat{\boldsymbol{\gamma}}_{\text{RR}} & =\sigma_{\epsilon}^{-2}\Lambda\tilde{Z}^{T}\left[\sigma_{\epsilon}^{-2}\tilde{Z}\Lambda\tilde{Z}^{T}+I\right]^{-1}\mathbf{y}\\
& =\Lambda\tilde{Z}^{T}\left[\tilde{Z}\Lambda\tilde{Z}^{T}+\Sigma\right]^{-1}\mathbf{y}\\
& =\hat{\boldsymbol{\gamma}}_{\text{BLUP}}
\end{align*}
This is a well known connection between the RR estimator and BLUP
which proves the following result:

\begin{prop}
	Assuming $\hat{\Sigma}_{0}\propto I$, the prediction for $\Gamma$
	obtained by the MrRCE algorithm is equivalent to the multivariate
	RR estimator with Ridge matrix $K=\left(\hat{\Sigma}_{0}\otimes I_{p}\right)\hat{\Lambda}^{-1}$.
\end{prop}

\noindent To better understand this result, consider the case $\Sigma_{0}=\sigma_{\epsilon}^{2}I_{q}$
and $\Lambda_{0}=\sigma_{\gamma}^{2}C$, where $C=C_{\rho}$ is an
equicorrelation matrix with parameter $\rho$. Let $K=\left(\Sigma_{0}\otimes I_{p}\right)\Lambda^{-1}=\eta C^{-1}\otimes I_{p}$
where $\eta=\left(\sigma_{\epsilon}/\sigma_{\gamma}\right)^{2}$.
It is easy to verify that $C^{-1}$ is itself an equicorrelation matrix,
$C^{-1}=aI_{q}+bJ_{q}$, where, 
\begin{align*}
a & =\frac{1}{1-\rho}\text{, }b=\frac{-\rho}{1-\rho}\left[\frac{1}{1+\left(q-1\right)\rho}\right]
\end{align*}
For simplicity, we only examine the penalty structure for $q=2,p=1$.
Denote the coefficients vector by $\boldsymbol{\gamma}=\left(\gamma_{11},\gamma_{12}\right)^{T}$.
The ridge penalty is given by, 
\begin{align}
\eta\left[\boldsymbol{\gamma}^{T}C^{-1}\boldsymbol{\gamma}\right] & =\eta\left[\left(a+b\right)\left\Vert \boldsymbol{\gamma}\right\Vert _{2}^{2}+2b\gamma_{11}\cdot\gamma_{12}\right]\label{eq:3.2}\\
& =\eta\frac{1}{1-\rho^{2}}\left\Vert \boldsymbol{\gamma}\right\Vert _{2}^{2}+2\eta b\left(\gamma_{11}\cdot\gamma_{12}\right)\nonumber 
\end{align}
Note that (\ref{eq:3.2}) can be reduced to the univariate ridge penalty
by setting $\rho=0$, i.e., by considering $i.i.d$ coefficients.
For $\rho>0$, the second term in (\ref{eq:3.2}) kicks-in. We note
that $b<0$ for $\rho\in\left(0,1\right)$, meaning that the second
penalty term in (\ref{eq:3.2}) is negative, for same sign coefficients.
This simple example illustrates that the MrRCE method favors equal
sign coefficients, within groups.

\section{\label{sec:sec4}Simulation Study}

In this section, we compare the performance of the MrRCE method to
other multivariate regression estimators, over several settings of
simulated data sets. We show that the MrRCE method significantly outperforms
all competitors, in terms of Model Error, for the vast majority of
simulated settings.

\subsection{Estimators}

We construct estimators using natural competitors of the MrRCE method,
and report the results for the following methods:
\begin{enumerate}
	\item \textit{Ordinary Least Squares (OLS)}: Perform $q$ separate LS regressions.
	\item \textit{Group Lasso}: Place an $L_{1}/L_{2}$-penalty over the rows
	of the coefficient matrix, with $3$-fold cross-validation (CV) for
	the selection the tuning parameter.
	\item \textit{Ridge Regression}: The tuning parameter is selected via leave-one-out
	cross-validation (LOO-CV) and is shared across all task.
	\item \textit{MRCE}: The tuning parameters are selected using $5$-fold
	CV.
	\item \textit{MrRCE}: The $L_{1}$-regularization parameter (for the graphical
	lasso algorithm) is selected via 3-fold CV.
\end{enumerate}

\subsection{Models}

For each settings and every replication, we generate an $n\times p$
predictor matrix $Z$ with rows drawn independently from $N_{p}\left(0,\Sigma_{Z}\right)$,
where $\left(\Sigma_{Z}\right)_{ij}=\rho_{Z}^{\left|i-j\right|}$
and $\rho_{Z}=.7$ (similar to \citet{yuan2007dimension,peng2010regularized,rothman2010sparse}).
Following \citet{rothman2010sparse}, the coefficient matrix $\Gamma$
is generated as the element-wise product of three matrices: First,
we sample a $p\times q$ matrix $W\sim MVN_{p\times q}\left(0,I_{p},\sigma^{2}C_{\rho}\right)$,
with $C_{\rho}=I+\rho\left(J-I\right)$, where $J$ is a matrix of
ones and $I$ is the identity matrix, both of dimensions $q\times q$.
The values of $\rho$ are ranging from $0$ to $0.8$, where $\rho=0$
corresponds to $i.i.d$ samples, $\gamma_{ij}\sim N\left(0,\sigma^{2}\right)$.
Next, we set, 
\[
\Gamma=W\odot K\odot Q
\]
where $\odot$ denotes the element-wise product. The entries of
the $p\times q$ matrix $K$ are drawn independently from $\text{Ber}\left(1-s\right)$,
and the elements in each row of the matrix $Q$ are all equal zero
or one, according to $p$ independent Bernoulli draws with success
probability $1-s_{g}$. Hence, setting $s,s_{g}>0$ will induce element-wise
and group sparsity in $\Gamma$. The rows of the error matrix $E$
are drawn independently from $N_{q}\left(0,\Sigma\right)$. We consider
several structures for the error covariance matrix, specified in the
form of the transformed error covariance matrix, $\tilde{\Sigma}:=U^{T}\Sigma U$,
where $U$ is the orthogonal matrix obtained via eigendecomposition
over the matrix $C_{\rho}$ (see Eq. \ref{eq:eigen}):
\begin{enumerate}
	\item \textit{Independent Errors.} The errors are drawn i.i.d form $N_{q}\left(0,I_{q}\right)$.
	\item \textit{Autoregressive Error Covariance \textemdash{} $AR\left(1\right)$}.
	We let $\tilde{\Sigma}_{ij}=\rho_{E}^{\left|i-j\right|}$. The transformed
	error covariance matrix is dense, whereas the precision matrix $\tilde{\Omega}$
	is a sparse, banded matrix.
	\item \textit{Fractional Gaussian Noise (FGN)}. The transformed error covariance
	matrix is given by, 
	\[
	\tilde{\Sigma}_{i,j}=.5\left[\left(\left|i-j\right|+1\right){}^{2H}-2\left|i-j\right|^{2H}+\left(\left|i-j\right|-1\right){}^{2H}\right]
	\]
	with $H=.95$. Both the transformed error covariance matrix $\tilde{\Sigma}$
	and its inverse have a dense structure.
	\item \textit{Equicorrelation Covariance Structure}. We let $\tilde{\Sigma}_{ij}=\rho_{E}$
	for $j\neq i$, and $\tilde{\Sigma}_{ij}=1$ for $j=i$. Both the
	transformed error covariance matrix and its inverse have a dense structure. 
\end{enumerate}

\subsection{Performance Measure}

For a given realization of the coefficient matrix and method $m$,
and for each replication $r$, let $\boldsymbol{\gamma}_{j}^{\left(r\right)}$
denote the true coefficient vector and $\hat{\boldsymbol{\gamma}}_{j}^{\left(r\right)}\left(m\right)$
denote the estimated coefficient vector, both for the $j$th response.
The mean-squared estimation error is given by,
\begin{align*}
ME_{m}^{\left(r\right)}\left(\boldsymbol{\gamma}_{j}^{\left(r\right)},\hat{\boldsymbol{\gamma}}_{j}^{\left(r\right)}\left(m\right)\right) & =\intop\left[\left(\boldsymbol{\gamma}_{j}^{\left(r\right)}-\hat{\boldsymbol{\gamma}}_{j}^{\left(r\right)}\left(m\right)\right)^{T}z\right]^{2}p\left(z\right)dz\\
& =\left(\boldsymbol{\gamma}_{j}^{\left(r\right)}-\hat{\boldsymbol{\gamma}}_{j}^{\left(r\right)}\left(m\right)\right)^{T}\Sigma_{Z}\left(\boldsymbol{\gamma}_{j}^{\left(r\right)}-\hat{\boldsymbol{\gamma}}_{j}^{\left(r\right)}\left(m\right)\right)
\end{align*}
where $p\left(z\right)$ and $\Sigma_{Z}$ are the density function
and covariance matrix of $z$, respectively. We evaluate the performance
using the model error (ME), following the approach of \citet{breiman1997predicting,yuan2007dimension,rothman2010sparse},
\[
ME_{m}^{\left(r\right)}\left(\Gamma^{\left(r\right)},\hat{\Gamma}^{\left(r\right)}\left(m\right)\right)=\text{tr}\left[\left(\Gamma^{\left(r\right)}-\hat{\Gamma}^{\left(r\right)}\left(m\right)\right)^{T}\Sigma_{Z}\left(\Gamma^{\left(r\right)}-\hat{\Gamma}^{\left(r\right)}\left(m\right)\right)\right]
\]
The ME over all $N$ replications is averaged to obtain our performance
measure,
\[
ME_{m}=\frac{1}{N}\sum_{r=1}^{N}ME_{m}^{\left(r\right)}
\]

\subsection{Results}

\begin{figure}
	\centering{}
	\includegraphics[scale=0.35]{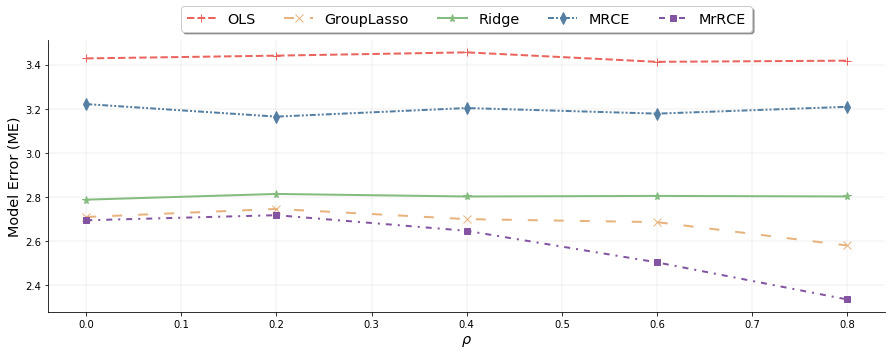}
	\caption{
		\label{fig:Id}\textit{Independent Errors}. Average model error (ME)
		versus the correlation parameter $\rho$, based on $N=200$ replications
		with $n=50,p=20,q=5$ and sparsity levels $s=0.2,s_{g}=0$.}
\end{figure}

\begin{figure}
	\centering{}
	\includegraphics[scale=0.35]{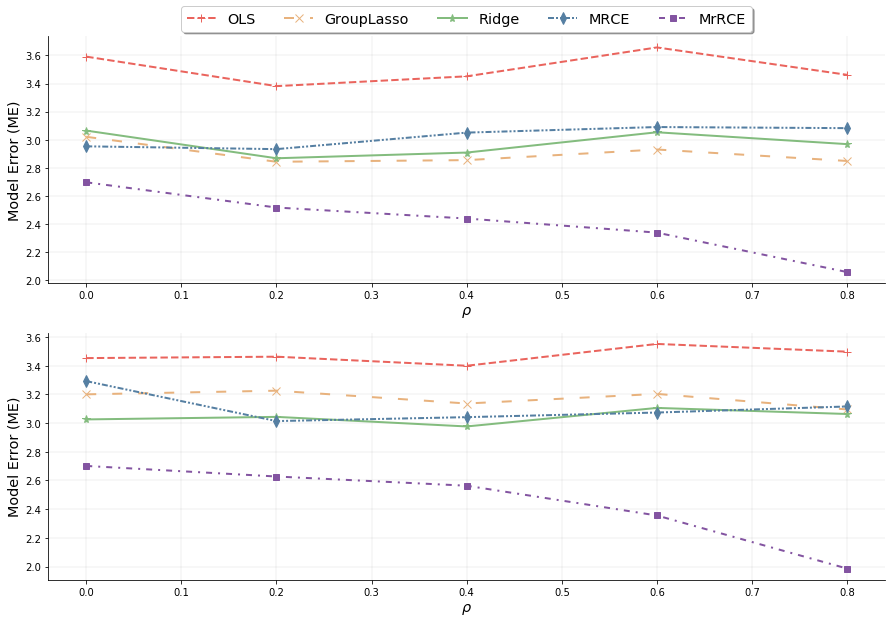}
	\caption{\label{fig:ar}\textit{Autoregressive}. Average model error (ME) versus
		the correlation parameter $\rho$, based on $N=200$ replications
		with $n=50,p=20,q=5$. Top: $s=s_{g}=0.1$. Bottom: $s=s_{g}=0$.}
\end{figure}

\begin{figure}
	\centering
	\includegraphics[scale=0.35]{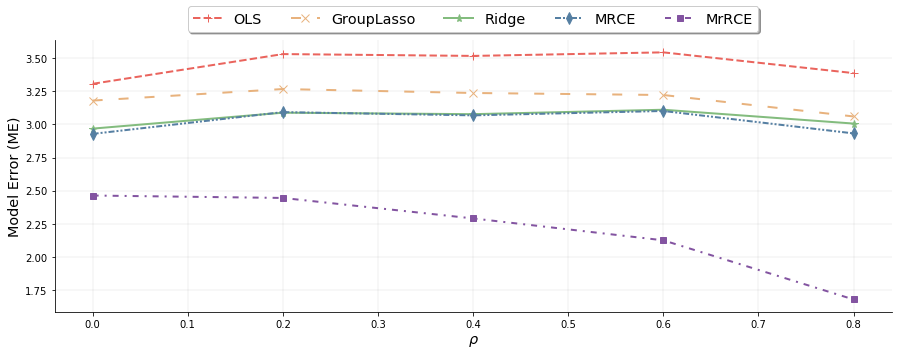}
	\caption{\label{fig:fgn}\textit{Fractional Gaussian Noise}. Average model
		error (ME) versus the correlation parameter $\rho$, based on $N=200$
		replications with $n=50,p=20,q=5$ and sparsity levels $s=s_{g}=0$. }
\end{figure}

\begin{figure}
	\centering
	\includegraphics[scale=0.35]{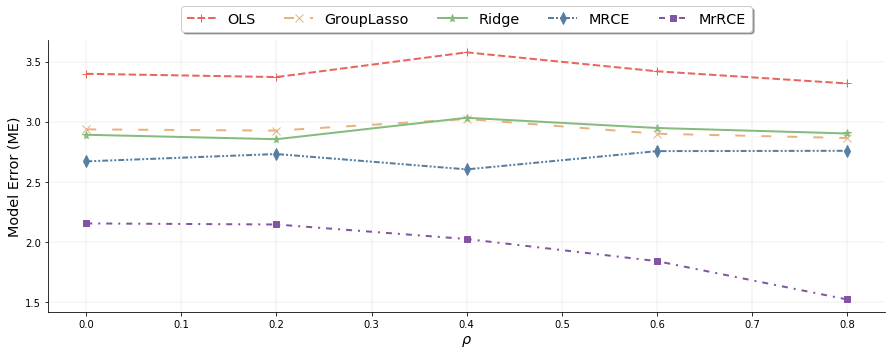}\caption{\label{fig:equic}\textit{Equicorrelation}. Average model error (ME)
		versus the correlation parameter $\rho$, based on $N=200$ replications
		with $n=50,p=20,q=5$ and sparsity levels $s=s_{g}=0.1$.}
\end{figure}

We simulate $N=200$ replications with $n=50$, $p=20$ and $q=5$,
for each setting. The correlation parameter $\rho$ ranges from $0$
to $0.8$, with $0.2$ steps. Significance tests were performed using
paired $t-$test.\textbf{\textit{}}\\
\textbf{\textit{}}\\
\textbf{\textit{Independent Errors}}. We first consider an identity
error covariance structure, $\tilde{\Sigma}=I_{q}$, and set the sparsity
and group sparsity levels at $s=0.2,s_{g}=0$. Hence, for small values
of $\rho$ we do not expect any advantage for our method over the
competitors. The average ME is displayed in Figure \ref{fig:Id}.
Indeed, for $\rho=0,.2$, our method achieves no significant improvement
over Group Lasso. For $\rho>.2$, the MrRCE method achieves significant
improvement over all competitors (all $p$-values $<1\text{e}-2$).
\\
\\
\textbf{\textit{Autoregressive (AR)}}. Let $\tilde{\Sigma_{ij}}=\rho_{E}^{\left|i-j\right|}$,
with $\rho_{E}=0.75$. We use two settings for the sparsity levels,
$s=s_{g}=0$, and $s=s_{g}=0.1$. Although the transformed precision
matrix is a sparse, banded matrix, the assumptions of MrRCE only partially
hold, as we induce sparsity in $\Gamma$ as well. The results are
displayed in Figure \ref{fig:ar}. For both settings, the MrRCE method
achieves the best ME performance, with a significant improvement over
competing methods (all $p$-values $<1\text{e}-3$).
\\
\\
\textbf{\textit{Fractional Gaussian Noise}}. This covariance structure
for the error terms was also considered by \citet{rothman2010sparse}.
We construct a dense coefficient matrix, by setting $s=s_{g}=0$.
The results are presented in Figure \ref{fig:fgn}, showing that our
proposed method provides a considerable improvement over competitors
(all $p$-values $<1\text{e}-19$). The margin by which MrRCE outperforms
the other methods increases with $\rho$. 
\\
\\
\textbf{\textit{Equicorrelation}}. Finally, we let $\tilde{\Sigma}_{ij}=\rho_{E}=0.9$
for $i\neq j$, and set $s=s_{g}=0.1$. The results are displayed
in Figure \ref{fig:equic}. The MRCE method exploits the correlated
errors, achieving better performance than the Group Lasso, Ridge and
OLS methods, and is second only to MrRCE, which significantly outperforms
all competitor methods for all values of $\rho$ (all $p$-values
$<1\text{e}-8$).

\section{\label{sec:sec5}Applications}

We consider two publicly available real-life datasets:
\begin{enumerate}
	\item \textit{NYC Taxi Rides}\footnote{The data is available at http://www.nyc.gov/html/tlc/html/about/trip\_record\_data.shtml.}.
	The data consists of the daily number of New-York City (NYC) taxi
	rides, ranging from January 2016 to December 2017.
	\item \textit{Avocado Prices}\footnote{The data is available at https://www.kaggle.com/neuromusic/avocado-prices.}.
	The data was provided by the Hass Avocado Board website and represents
	weekly retail scan data for national retail volume (units) and price.
\end{enumerate}
We measure and report the performance of the following methods:
\begin{enumerate}
	\item \textit{Ordinary Least Squares}.
	\item \textit{Group Lasso}. Apply $3$-fold CV for the selection of the
	tuning parameter.
	\item \textit{Separate Lasso}. Perform $q$ separate lasso regression models
	with $3$-fold CV for selecting the tuning parameters.
	\item \textit{Ridge Regression}. Perform $q$ separate ridge regression
	models, with shared regularization parameter, selected via LOO-CV
	(e.g. same ridge penalty for all $pq$ parameters).
	\item \textit{Separate Ridge Regression}. Perform $q$ separate ridge regression
	models with LOO-CV for selecting the tuning parameters.
	\item \textit{MRCE}. Apply 5-fold CV for selecting the regularization parameters.
	\item \textit{MrRCE}. Apply $3$-fold CV for selecting the graphical lasso
	regularization parameter.\\
\end{enumerate}
\textbf{\textit{NYC Taxi Rides}}. We consider the problem of forecasting
the performance of $q=2$ taxi vendors in NYC, using historical records
of the daily number of rides, spanning from January 2016 to December
2017 ($n=730$). This multivariate time-series data is generated according
to human activities and actions, and as such can be expected to be
strongly affected by multiple seasonalities and holidays effects.
For a regular period $P$, we utilize the Fourier series to model
the periodic effects \citep{HARVEY1993261,taylor2018forecasting},
by constructing $2\cdot N_{P}$ features of the form, 
\[
Z_{P}\left(t\right)=\left\{ \cos\left(\frac{2\pi nt}{P}\right),\sin\left(\frac{2\pi nt}{P}\right)\right\} _{n=1,...,N_{P}}
\]
We account for the weekly and yearly seasonalities and introduce the
corresponding $P$-cyclic covariates. For a holiday $H$, which occurs
at times $T\left(H\right)$, we use a simple indicator predictors
of the form, 
\[
Z_{H}\left(t\right)=\mathbbm1_{\left\{ t\in T\left(H\right)\right\} }
\]
Lastly, we incorporate covariates for the modeling of a piecewise
linear trend. These transformations shift the multivariate time-series
problem into a feature space with $p=68$, where the linear assumption
is appropriate. We denote the transformed observations by,
\[
\left\{ Z\left(t\right),Y\left(t\right)\right\} _{t=1,...,T}
\]
where $Z\left(t\right)\in\mathbb{R}^{p}$ contains measurements of
the covariates, $Y\left(t\right)\in\mathbb{R}^{q}$ contains the $q$
responses, and $Y_{j}\left(t\right)\in\left[0,1\right]$ represents
the scaled response of the $j$th task at time $t$, obtained by dividing
the original observation by the maximal response value for that given
task.

We evaluate the forecast performance of the different methods using
cross-validation like approach, in which we produce $K$ forecasts
at multiple cutoff points along the history \citep{taylor2018forecasting}.
For cutoff $k=0,...,K-1$, we use the first $n_{train,k}=365+k\cdot14$
days for training, and the next $n_{test}=14$ observations as the
test set. The performance of method $m$ over the $k$th ``fold''
is measured according to the Mean Squared Error (MSE), 
\[
\text{MSE}_{k}^{m}=\frac{1}{n_{test}}\cdot\frac{1}{q}\sum_{t\in T_{k}}\sum_{j=1}^{q}\left(y_{j,t}-\hat{y}_{j,t}\left(m\right)\right)^{2}
\]
where $T_{k}$ are the time indices for the $k$th test set, and $\hat{y}_{j,t}\left(m\right)$
is the forecast for the $j$th task at time $t$, produced using method
$m$. Using the above procedure, we obtain $K=26$ realizations of
the MSE, $\left\{ MSE_{k}^{m}\right\} _{k=0}^{K-1}$, for each method
$m$. The mean and standard deviation of the MSE for each of the methods
are reported in Table \ref{tab:nyc-res}. The MrRCE method
attains the best forecast performance, with lowest mean MSE and smallest
standard deviation, followed by the Ridge and Separate-Ridge methods.
A paired $t$-test confirms that the improvement in accuracy achieved
by our method is significant (all $p$-values $<0.05$). We also note
that the estimated similarity level for this data is $\hat{\rho}=0.992$.

\begin{table}
	
	\caption{\label{tab:nyc-res}\textit{NYC Taxi Rides}. Mean and standard deviation of the MSE, estimated over $K=26$ cutoffs.}
	
	\centering
	
	\begin{tabular*}{8cm}{@{\extracolsep{\fill}}lcc}
		\toprule 
		\textbf{Model} & \textbf{Mean} & \textbf{Std}\tabularnewline
		\midrule
		MrRCE & \textbf{3.85e-3} & 4.57e-3\tabularnewline
		
		Ridge & 4.59e-3 & 5.34e-3\tabularnewline
		
		Sep. Ridge & 4.59e-3 & 5.34e-3\tabularnewline
		
		MRCE & 4.61e-3 & 5.12e-3\tabularnewline
		
		Group Lasso & 5.68e-3 & 7.72e-3\tabularnewline
		
		Sep. Lasso & 5.75e-3 & 7.12e-3\tabularnewline
		
		OLS & 2.00e-2 & 1.40e-2\tabularnewline
		
	\end{tabular*}

\end{table}

\textbf{\textit{Avocado Prices}}. We consider the weekly average avocado
prices for $q=5$ regions in the US, spanning from January 2015 to
April 2018 ($n=169$). We use national volume metrics and one hot
encoding for years ($p=12$) to predict the average avocado prices
for each region. The performance is measured according to the MSE,
with 10-fold CV. The mean and standard deviation of the MSE, calculated
over all folds, are reported in Table \ref{tab:avocado-res}. Our
proposed method attains the best prediction performance,
with lowest mean MSE and smallest standard deviation. A paired $t$-test
confirms that the improvement in accuracy is significant (all $p$-values
$<0.05$). We also report the estimated similarity level for this
data, at $\hat{\rho}=0.689$.

\begin{table}
	
	\caption{\label{tab:avocado-res}\textit{Avocado Prices}. Mean and standard deviation of the MSE, estimated over $K=10$ folds.}
	
	\centering
	
	\begin{tabular*}{8cm}{@{\extracolsep{\fill}}lcc}
		\toprule 
		\textbf{Model} & \textbf{Mean} & \textbf{Std}\tabularnewline
		\midrule 
		MrRCE & \textbf{53.9e-2} & 22.6e-2\tabularnewline
		
		MRCE & 63.4e-2 & 29.0e-2\tabularnewline
		
		Group Lasso & 66.7e-2 & 29.9e-2\tabularnewline
		
		Sep. Ridge & 71.0e-2 & 38.7e-2\tabularnewline
		
		Ridge & 71.5e-2 & 39.8e-2\tabularnewline
		
		Sep. Lasso & 72.0e-2 & 36.0e-2\tabularnewline
		
		OLS & 73.1e-2 & 41.3e-2\tabularnewline
		
	\end{tabular*}
	
\end{table}

\section{\label{sec:sec6}Summary and Discussion}

We have presented the MrRCE method to produce an estimator of the
covariance components and a predictor of the multivariate regression
coefficient matrix. Our method exploits similarities among random
coefficients and accounts for correlated errors. We have proposed
an efficient EM-based algorithm for computing MrRCE. By using simulated
and real data, we have illustrated that the proposed method can outperform
the commonly used methods for multivariate regression, in settings
were errors or coefficients are related. 

Our method can be extended in several ways. For example, one could
consider an arbitrary group structure over the coefficient matrix,
model the similarities via different covariance structure, or allow
for per-group similarity coefficient. In addition, one could extend
the MrRCE formulation to also allow for fixed effects, as in (\ref{eq:1.4}).

\section{Acknowledgement}

This research was partially supported by Israeli Science Foundation
grant 1804/16.

\printbibliography

@Article{breiman1997predicting,
  author    = {Breiman, Leo and Friedman, Jerome H},
  title     = {Predicting multivariate responses in multiple linear regression},
  journal   = {Journal of the Royal Statistical Society: Series B (Statistical Methodology)},
  year      = {1997},
  volume    = {59},
  number    = {1},
  pages     = {3--54},
  publisher = {Wiley Online Library},
}

@Article{rothman2010sparse,
  author    = {Rothman, Adam J and Levina, Elizaveta and Zhu, Ji},
  title     = {Sparse multivariate regression with covariance estimation},
  journal   = {Journal of Computational and Graphical Statistics},
  year      = {2010},
  volume    = {19},
  number    = {4},
  pages     = {947--962},
  publisher = {Taylor \& Francis},
}

@Article{henderson1959estimation,
  author    = {Henderson, Charles R and Kempthorne, Oscar and Searle, Shayle R and Von Krosigk, CM},
  title     = {The estimation of environmental and genetic trends from records subject to culling},
  journal   = {Biometrics},
  year      = {1959},
  volume    = {15},
  number    = {2},
  pages     = {192--218},
  publisher = {JSTOR},
}

@Article{wilms2018algorithm,
  author    = {Wilms, Ines and Croux, Christophe},
  title     = {An algorithm for the multivariate group lasso with covariance estimation},
  journal   = {Journal of Applied Statistics},
  year      = {2018},
  volume    = {45},
  number    = {4},
  pages     = {668--681},
  publisher = {Taylor \& Francis},
}

@Article{li2015multivariate,
  author    = {Li, Yanming and Nan, Bin and Zhu, Ji},
  title     = {Multivariate sparse group lasso for the multivariate multiple linear regression with an arbitrary group structure},
  journal   = {Biometrics},
  year      = {2015},
  volume    = {71},
  number    = {2},
  pages     = {354--363},
  publisher = {Wiley Online Library},
}

@Article{henderson1975best,
  author    = {Henderson, Charles R},
  title     = {Best linear unbiased estimation and prediction under a selection model},
  journal   = {Biometrics},
  year      = {1975},
  pages     = {423--447},
  publisher = {JSTOR},
}

@Article{henderson1984applications,
  author  = {Henderson, Charles R},
  title   = {Applications of linear models in animal breeding: University of Guelph},
  journal = {Applications of linear models in animal breeding, University of Guelph.},
  year    = {1984},
}

@Article{korte2012mixed,
  author    = {Korte, Arthur and Vilhj{\'a}lmsson, Bjarni J and Segura, Vincent and Platt, Alexander and Long, Quan and Nordborg, Magnus},
  title     = {A mixed-model approach for genome-wide association studies of correlated traits in structured populations},
  journal   = {Nature genetics},
  year      = {2012},
  volume    = {44},
  number    = {9},
  pages     = {1066},
  publisher = {Nature Publishing Group},
}

@Article{zhou2014efficient,
  author    = {Zhou, Xiang and Stephens, Matthew},
  title     = {Efficient multivariate linear mixed model algorithms for genome-wide association studies},
  journal   = {Nature methods},
  year      = {2014},
  volume    = {11},
  number    = {4},
  pages     = {407},
  publisher = {Nature Publishing Group},
}

@Article{furlotte2015efficient,
  author    = {Furlotte, Nicholas A and Eskin, Eleazar},
  title     = {Efficient multiple-trait association and estimation of genetic correlation using the matrix-variate linear mixed model},
  journal   = {Genetics},
  year      = {2015},
  volume    = {200},
  number    = {1},
  pages     = {59--68},
  publisher = {Genetics Soc America},
}

@Article{friedman2008sparse,
  author    = {Friedman, Jerome H and Hastie, Trevor and Tibshirani, Robert},
  title     = {Sparse inverse covariance estimation with the graphical lasso},
  journal   = {Biostatistics},
  year      = {2008},
  volume    = {9},
  number    = {3},
  pages     = {432--441},
  publisher = {Oxford University Press},
}

@Article{yuan2007dimension,
  author    = {Yuan, Ming and Ekici, Ali and Lu, Zhaosong and Monteiro, Renato},
  title     = {Dimension reduction and coefficient estimation in multivariate linear regression},
  journal   = {Journal of the Royal Statistical Society: Series B (Statistical Methodology)},
  year      = {2007},
  volume    = {69},
  number    = {3},
  pages     = {329--346},
  publisher = {Wiley Online Library},
}

@Article{banerjee2008model,
  author  = {d{\'A}spremont, Alexandre and Banerjee, Onureena and El Ghaoui, Laurent},
  title   = {Model selection through sparse maximum likelihood estimation for multivariate gaussian or binary data},
  journal = {Journal of Machine learning research},
  year    = {2008},
  volume  = {9},
  number  = {Mar},
  pages   = {485--516},
}

@Article{yuan2006model,
  author    = {Yuan, Ming and Lin, Yi},
  title     = {Model selection and estimation in regression with grouped variables},
  journal   = {Journal of the Royal Statistical Society: Series B (Statistical Methodology)},
  year      = {2006},
  volume    = {68},
  number    = {1},
  pages     = {49--67},
  publisher = {Wiley Online Library},
}

@Article{rothman2008sparse,
  author    = {Rothman, Adam J and Bickel, Peter J and Levina, Elizaveta and Zhu, Ji},
  title     = {Sparse permutation invariant covariance estimation},
  journal   = {Electronic Journal of Statistics},
  year      = {2008},
  volume    = {2},
  pages     = {494--515},
  publisher = {The Institute of Mathematical Statistics and the Bernoulli Society},
}

@InProceedings{sohn2012joint,
  author    = {Sohn, Kyung-Ah and Kim, Seyoung},
  title     = {Joint estimation of structured sparsity and output structure in multiple-output regression via inverse-covariance regularization},
  booktitle = {Artificial Intelligence and Statistics},
  year      = {2012},
  pages     = {1081--1089},
}

@InProceedings{hsieh2011sparse,
  author    = {Hsieh, Cho-Jui and Dhillon, Inderjit S and Ravikumar, Pradeep K and Sustik, M{\'a}ty{\'a}s A},
  title     = {Sparse inverse covariance matrix estimation using quadratic approximation},
  booktitle = {Advances in neural information processing systems},
  year      = {2011},
  pages     = {2330--2338},
}

@Article{peng2010regularized,
  author    = {Peng, Jie and Zhu, Ji and Bergamaschi, Anna and Han, Wonshik and Noh, Dong-Young and Pollack, Jonathan R and Wang, Pei},
  title     = {Regularized multivariate regression for identifying master predictors with application to integrative genomics study of breast cancer},
  journal   = {The annals of applied statistics},
  year      = {2010},
  volume    = {4},
  number    = {1},
  pages     = {53},
  publisher = {NIH Public Access},
}

@Article{tibshirani1996regression,
  author    = {Tibshirani, Robert},
  title     = {Regression shrinkage and selection via the lasso},
  journal   = {Journal of the Royal Statistical Society. Series B (Methodological)},
  year      = {1996},
  pages     = {267--288},
  publisher = {JSTOR},
}

@Book{gupta2018matrix,
  title     = {Matrix variate distributions},
  publisher = {Chapman and Hall/CRC},
  year      = {2018},
  author    = {Gupta, Arjun K and Nagar, Daya K},
}

@Article{hoerl1970ridge,
  author    = {Hoerl, Arthur E and Kennard, Robert W},
  title     = {Ridge regression: Biased estimation for nonorthogonal problems},
  journal   = {Technometrics},
  year      = {1970},
  volume    = {12},
  number    = {1},
  pages     = {55--67},
  publisher = {Taylor \& Francis Group},
}

@Article{chen2016sparse,
  author    = {Chen, Lisha and Huang, Jianhua Z},
  title     = {Sparse reduced-rank regression with covariance estimation},
  journal   = {Statistics and Computing},
  year      = {2016},
  volume    = {26},
  number    = {1-2},
  pages     = {461--470},
  publisher = {Springer},
}

@InProceedings{obozinski2009high,
  author    = {Obozinski, Guillaume R and Wainwright, Martin J and Jordan, Michael I},
  title     = {High-dimensional support union recovery in multivariate regression},
  booktitle = {Advances in Neural Information Processing Systems},
  year      = {2009},
  pages     = {1217--1224},
}

@Article{kim2012tree,
  author    = {Kim, Seyoung and Xing, Eric P},
  title     = {Tree-guided group lasso for multi-response regression with structured sparsity, with an application to eQTL mapping},
  journal   = {The Annals of Applied Statistics},
  year      = {2012},
  volume    = {6},
  number    = {3},
  pages     = {1095--1117},
  publisher = {Institute of Mathematical Statistics},
}

@Article{anderson1951estimating,
  author    = {Anderson, Theodore Wilbur},
  title     = {Estimating linear restrictions on regression coefficients for multivariate normal distributions},
  journal   = {The Annals of Mathematical Statistics},
  year      = {1951},
  pages     = {327--351},
  publisher = {JSTOR},
}

@Article{izenman1975reduced,
  author    = {Izenman, Alan Julian},
  title     = {Reduced-rank regression for the multivariate linear model},
  journal   = {Journal of multivariate analysis},
  year      = {1975},
  volume    = {5},
  number    = {2},
  pages     = {248--264},
  publisher = {Elsevier},
}

@Book{velu2013multivariate,
  title     = {Multivariate reduced-rank regression: theory and applications},
  publisher = {Springer Science \& Business Media},
  year      = {2013},
  author    = {Velu, Raja and Reinsel, Gregory C},
  volume    = {136},
}

@Article{turlach2005simultaneous,
  author    = {Turlach, Berwin A and Venables, William N and Wright, Stephen J},
  title     = {Simultaneous variable selection},
  journal   = {Technometrics},
  year      = {2005},
  volume    = {47},
  number    = {3},
  pages     = {349--363},
  publisher = {Taylor \& Francis},
}

@Article{lee2012simultaneous,
  author    = {Lee, Wonyul and Liu, Yufeng},
  title     = {Simultaneous multiple response regression and inverse covariance matrix estimation via penalized Gaussian maximum likelihood},
  journal   = {Journal of multivariate analysis},
  year      = {2012},
  volume    = {111},
  pages     = {241--255},
  publisher = {Elsevier},
}

@Article{dawid1981some,
  author    = {Dawid, Philip A},
  title     = {Some matrix-variate distribution theory: notational considerations and a Bayesian application},
  journal   = {Biometrika},
  year      = {1981},
  volume    = {68},
  number    = {1},
  pages     = {265--274},
  publisher = {Oxford University Press},
}

@Article{vattikuti2012heritability,
  author    = {Vattikuti, Shashaank and Guo, Juen and Chow, Carson C},
  title     = {Heritability and genetic correlations explained by common SNPs for metabolic syndrome traits},
  journal   = {PLoS genetics},
  year      = {2012},
  volume    = {8},
  number    = {3},
  pages     = {e1002637},
  publisher = {Public Library of Science},
}

@Article{kruuk2004estimating,
  author    = {Kruuk, Loeske EB},
  title     = {Estimating genetic parameters in natural populations using the 'animal model'},
  journal   = {Philosophical Transactions of the Royal Society of London B: Biological Sciences},
  year      = {2004},
  volume    = {359},
  number    = {1446},
  pages     = {873--890},
  publisher = {The Royal Society},
}

@Article{kang2010variance,
  author    = {Kang, Hyun Min and Sul, Jae Hoon and Zaitlen, Noah A and Kong, Sit-yee and Freimer, Nelson B and Sabatti, Chiara and Eskin, Eleazar and others},
  title     = {Variance component model to account for sample structure in genome-wide association studies},
  journal   = {Nature genetics},
  year      = {2010},
  volume    = {42},
  number    = {4},
  pages     = {348},
  publisher = {Nature Publishing Group},
}

@Article{bunea2011optimal,
  author    = {Bunea, Florentina and She, Yiyuan and Wegkamp, Marten H},
  title     = {Optimal selection of reduced rank estimators of high-dimensional matrices},
  journal   = {The Annals of Statistics},
  year      = {2011},
  pages     = {1282--1309},
  publisher = {JSTOR},
}

@Article{deshpande2017simultaneous,
  author  = {Deshpande, Sameer K and Rockova, Veronika and George, Edward I},
  title   = {Simultaneous Variable and Covariance Selection with the Multivariate Spike-and-Slab Lasso},
  journal = {arXiv preprint arXiv:1708.08911},
  year    = {2017},
}

@Article{witten2009covariance,
  author    = {Witten, Daniela M and Tibshirani, Robert},
  title     = {Covariance-regularized regression and classification for high dimensional problems},
  journal   = {Journal of the Royal Statistical Society: Series B (Statistical Methodology)},
  year      = {2009},
  volume    = {71},
  number    = {3},
  pages     = {615--636},
  publisher = {Wiley Online Library},
}

@Article{sherman1950adjustment,
  author    = {Sherman, Jack and Morrison, Winifred J},
  title     = {Adjustment of an inverse matrix corresponding to a change in one element of a given matrix},
  journal   = {The Annals of Mathematical Statistics},
  year      = {1950},
  volume    = {21},
  number    = {1},
  pages     = {124--127},
  publisher = {JSTOR},
}

@Article{woodbury1950inverting,
  author  = {Woodbury, Max A},
  title   = {Inverting modified matrices},
  journal = {Memorandum report},
  year    = {1950},
  volume  = {42},
  number  = {106},
  pages   = {336},
}

@Article{brown1980adaptive,
  author    = {Brown, Philip J and Zidek, James V},
  title     = {Adaptive multivariate ridge regression},
  journal   = {The Annals of Statistics},
  year      = {1980},
  pages     = {64--74},
  publisher = {JSTOR},
}

@InCollection{HARVEY1993261,
  author    = {Andrew, Harvey C and Neil, Shephard},
  title     = {Structural time series models},
  booktitle = {Econometrics},
  publisher = {Elsevier},
  year      = {1993},
  volume    = {11},
  series    = {Handbook of Statistics},
  pages     = {261 - 302},
  issn      = {0169-7161},
}

@Article{taylor2018forecasting,
  author    = {Taylor, Sean J and Letham, Benjamin},
  title     = {Forecasting at scale},
  journal   = {The American Statistician},
  year      = {2018},
  volume    = {72},
  number    = {1},
  pages     = {37--45},
  publisher = {Taylor \& Francis},
}

@Article{yuan2007model,
  author    = {Yuan, Ming and Lin, Yi},
  title     = {Model selection and estimation in the Gaussian graphical model},
  journal   = {Biometrika},
  year      = {2007},
  volume    = {94},
  number    = {1},
  pages     = {19--35},
  publisher = {Oxford University Press},
}

@article{yin2011sparse,
  title={A sparse conditional gaussian graphical model for analysis of genetical genomics data},
  author={Yin, Jianxin and Li, Hongzhe},
  journal={The annals of applied statistics},
  volume={5},
  number={4},
  pages={2630},
  year={2011},
  publisher={NIH Public Access}
}

@book{chatfield2010introduction,
  title={An introduction to generalized linear models},
  author={Chatfield, Chris and Zidek, Jim and Lindsey, Jim},
  year={2010},
  publisher={Chapman and Hall/CRC}
}

@article{green1990use,
  title={On use of the EM algorithm for penalized likelihood estimation},
  author={Green, Peter J},
  journal={Journal of the Royal Statistical Society: Series B (Methodological)},
  volume={52},
  number={3},
  pages={443--452},
  year={1990},
  publisher={Wiley Online Library}
}

@article{dempster1977maximum,
  title={Maximum likelihood from incomplete data via the EM algorithm},
  author={Dempster, Arthur P and Laird, Nan M and Rubin, Donald B},
  journal={Journal of the Royal Statistical Society: Series B (Methodological)},
  volume={39},
  number={1},
  pages={1--22},
  year={1977},
  publisher={Wiley Online Library}
}

\end{document}